\documentclass[runningheads]{llncs}
\usepackage[T1]{fontenc}
\usepackage{subcaption}
\usepackage{graphicx}
\usepackage{wrapfig,lipsum,booktabs}
\usepackage{hyperref}

\usepackage{graphicx}
\usepackage{hyperref}
\usepackage{cite}
\usepackage{xspace}
\usepackage{xparse}

\usepackage{color}

\newcommand{\Algoname}{PrivComp-KG\xspace}

\begin{document}
\title{\Algoname: Leveraging Knowledge Graph and Large Language Models for Privacy Policy Compliance Verification}
\author{Leon Garza\inst{1} \and
Lavanya Elluri\inst{2} \and
Anantaa Kotal\inst{3} \and 
Aritran Piplai\inst{1} \and
Deepti Gupta\inst{2} \and
Anupam Joshi\inst{2}
}
\authorrunning{L. Garza et al.}
\institute{University of Texas at El Paso, 
\email{lgarza3@miners.utep.edu, apiplai@utep.edu}\and
Texas A\&M University Central Texas, 
\email{\{elluri,d.gupta\}@tamuct.edu}\\ \and
University of Maryland Baltimore County,
\email{\{anantak1,joshi\}@umbc.edu}\\
}

\titlerunning{PrivComp-KG: Leveraging KG and LLM for Compliance Verification}
\maketitle              % typeset the header of the contribution

\begin{abstract}
Data protection and privacy is becoming increasingly crucial in the digital era. Numerous companies depend on third-party vendors and service providers to carry out critical functions within their operations, encompassing tasks such as data handling and storage. However, this reliance introduces potential vulnerabilities, as these vendors' security measures and practices may not always align with the standards expected by regulatory bodies. In response, federal and international organizations have enacted data protection laws, regulations, and guidelines that govern vendor management and data dissemination. Businesses are required, often under the penalty of law, to ensure compliance with the evolving regulatory rules. However, interpreting and implementing these regulations pose challenges due to their complexity. Regulatory documents are extensive, demanding significant effort for interpretation, while vendor-drafted privacy policies often lack the detail required for full legal compliance, leading to ambiguity. To ensure a concise interpretation of the regulatory requirements and compliance of organizational privacy policy with said regulations, we propose a Large Language Model (LLM) and Semantic Web based approach for privacy compliance. In this paper, we develop the novel Privacy Policy Compliance Verification Knowledge Graph, \Algoname. The \Algoname is designed to efficiently store and retrieve comprehensive information concerning privacy policies, regulatory frameworks, and domain-specific knowledge pertaining to the legal landscape of privacy. Using LLMs and Retrieval Augmented Generation, we identify the relevant sections in a privacy policy with corresponding regulatory rules. Our LLM-based retrieval system achieved a correctness score of 0.9. This information about individual privacy policies is populated into the \Algoname. Combining this with the domain context and rules, the \Algoname can be queried to check for compliance with privacy policies by each vendor against relevant policy regulations. We demonstrate the relevance of the \Algoname, by verifying compliance of privacy policy documents for various organizations. This approach allows the policy writers to meaningfully understand the law's requirements, identify gaps in their existing policies, and update based on the evolving regulations. 

\end{abstract}
\section{Introduction}
%Every day large amounts of data is collected from various sources worldwide. 
Every day, vast quantities of data are collected from diverse sources across the globe. This spans from individual interactions on social media to industrial sensor readings. These collections of data are essential as it provides valuable insights into consumer behaviors, market trends, and societal patterns. Businesses leverage this data to tailor their products and services, optimize operations, and gain a competitive edge in the market. Researchers analyze vast datasets to make scientific breakthroughs, improve healthcare outcomes, and address societal challenges. This evolution coincides with a pervasive integration of technology into our daily lives, resulting in the widespread collection of individual and aggregated data.

However, the surge in data collection has sparked concerns regarding data privac. There is potential for misuse and unauthorized dissemination of consumers' private information by organizations collecting their data. Consequently, numerous data protection regulations, such as the European Union's General Data Protection Regulation (GDPR) \cite{gdpr}, Payment Card Industry Data Security Standard (PCI DSS) \cite{PCI_DSS}, and the California Consumer Privacy Protection Act (CCPA) \cite{CCPA,CPRA}, have been established in response to public apprehension. Privacy Policy regulations impose strict rules on collecting, using, and managing personal data. These rules require companies to follow principles like minimizing data collection, limiting its use to specified purposes, obtaining user consent, and ensuring the data accuracy, security, and the company's accountability. The GDPR, for example, sets a thorough set of privacy and data protection rules for companies accessing the European Union (EU) users' data. Known for setting some of the strictest data privacy and security standards worldwide, the GDPR is a model for properly using personal data, focusing on safeguarding and legally processing individuals' information. As a result, businesses must reassess how they handle data, ensuring their approaches comply with GDPR guidelines to protect people's privacy while maximizing benefits with data in this digital era. Furthermore, not adhering to this regulations not only makes the organization more susceptible to data breaches, but also holds them liable to pay huge penalties. 

In May 2023, the Irish Data Protection Commission (DPC) made a major move in the history of the GDPR by fining the American tech company Meta a record €1.2 billion \cite{MetaHitw39:online}. This fine was the highest ever because Meta moved the personal data of European users to the United States without ensuring enough protection for this data. This step by the DPC is a crucial moment in data protection law, highlighting how seriously rules about international data transfer are enforced under the GDPR. In July 2021, Amazon Europe Core SARL was fined €746 million by the Luxembourg National Commission for Data Protection (CNDP), marking the most significant penalty for violating the GDPR. This action followed a complaint from 10,000 individuals, organized by the French privacy group La Quadrature du Net, concerning Amazon's data processing practices. The CNDP's investigation revealed that Amazon's advertising targeting system operated without obtaining proper consent, contravening GDPR's stringent consent requirements, which demand clear communication and detailed explanations of personal data use, purpose, and usage. \cite{Amazon:online} In September 2023, TikTok faced a significant penalty from the Irish DPC, receiving a fine of EUR 345 million. This event marked one of the most considerable GDPR fines imposed on a social media platform, particularly highlighting issues around protecting children's data privacy. The decision underscored the critical need for tech companies to prioritize the safety of young users online. \cite{TikTok:oline} These instances underscores the critical need for businesses to understand privacy policy regulations and ensure that their data privacy policies are consistent with the regulations. 

Businesses must thoroughly understand the nature, scope, and purpose of the data they collect, process, and store. Furthermore, this information must be precisely recorded in a privacy policy document that's easy for users to access and understand. Writing a comprehensive privacy policy document is critical to building trust between data collectors and consumers. Developing a privacy policy that meets the extensive requirements of policy regulations is a significant challenge for companies, primarily because of the complexity of the regulation's rules. Privacy policies are short and concise for user ease while complying with all relevant sections of the regulatory document. Furthermore, the legal landscape of data protection regulations is dynamic and evolving. The GDPR sets a high privacy and data protection standard, yet it's subject to interpretation and ongoing adjustments by regulatory authorities. This ever-changing landscape necessitates that companies stay flexible and constantly monitor legal updates to ensure their privacy policies and practices align with compliance requirements. Additionally, the regulation applies to businesses operating within the EU and those outside the region if they process data from EU residents. This global reach complicates compliance efforts. Companies must navigate GDPR and any other local privacy laws that might apply. This complicates efforts to develop compliant privacy policies that satisfy all regulatory requirements. The motivation of this work is to help privacy policy writers efficiently develop a privacy policy in compliance with regulatory documents. 

We propose an innovative framework that enables policy writers to identify relevant regulatory rules, detect shortcomings in existing policies, and effectively address compliance requirements. In this work, we develop a Privacy Policy Compliance Verification Knowledge Graph, \Algoname, that is designed to collect and maintain information regarding policy and regulatory documents, and encapsulate domain knowledge. The inference rule engine is used to reason over the privacy policies and regulatory documents to verify compliance. The query engine is utilised to effectively gain this insight from the \Algoname and can be utilised by policy writers to identify any gaps in their existing policies. To efficiently populate the \Algoname with privacy policy documents, as well as their relevance to regulatory sections, we use Retrieval Augmented Generation (RAG) to assess privacy policies' alignment with GDPR articles. This avoids the need for constant model fine-tuning with evolving privacy laws. RAG helps generate responses by utilizing chunks of GDPR articles, allowing us to identify specific segments that match privacy policies, ensuring a dynamic and comprehensive approach without requiring continual model updates for each policy change. We demonstrate the utility of the \Algoname, by verifying compliance of privacy policy documents for various organizations in the OPP-115 dataset \cite{wilson2016creation}.

The structure of the paper is organized as follows. In Section \ref{Background} we describe relevant work in contemporary GDPR compliance efforts and LLM RAG methods for text retrieval. In Section \ref{Methodology}, we describe our framework leveraging LLMs for RAG to align GDPR sections with privacy policies. This section also discusses the development of \Algoname and illustrates how data retrieved from vendor policies are incorporated into the knowledge graph. In Section \ref{exp_res} we detail the utilisation of our framework to verify compliance of privacy policies in the OPP-115 dataset. Furthermore we describe the results from evaluation of the knowledge extraction and KG creation methods. The final section summarizes these discoveries and outlines prospective research avenues.

\section{Background and Related Work}
\label{Background}

\subsection{Privacy Policy Compliance}
As data collection increases, public concern over privacy risks has intensified, leading to the implementation of stringent privacy regulations like the GDPR. These regulations are crucial for safeguarding privacy, yet ensuring automatic compliance remains a complex challenge. The GDPR marks a significant change in privacy laws, setting up strict rules to protect people's rights and how companies must handle personal data. The European Union introduced it, requiring strict adherence from any organization within the EU or dealing with data from EU citizens. Ignoring these rules can lead to heavy fines of up to 20 million Euros or 4\% of a company's worldwide yearly income, depending on the seriousness of the violation. The GDPR applies to all organizations handling personal data, regardless of location, and replaces the older EU Data Protection Directive introduced in 1995, creating a unified privacy law for all EU countries. This means companies can now follow one set of privacy rules across the EU.\cite{gdpr} Furthermore, the GDPR's influence has reached beyond Europe, inspiring similar laws in other countries, including the United States. Given this move towards tighter privacy regulations globally, understanding and complying with the GDPR is crucial for software companies to meet current rules and prepare for any new privacy laws in the future.

In recent research \cite{wang2019data}, the concept of "Data Capsule" is introduced, automating compliance checks against privacy regulations in data processing. Individual data is associated with specific policies, ensuring adherence through residual policies and a new algorithm for effective policy derivation. This system advances individual privacy protection, albeit focusing solely on data subject rights. Another study \cite{liu2021have} proposes an approach to assess privacy policy alignment with GDPR Article 13 standards. By manually selecting 304 policies and developing a labeling system, the authors identify compliance issues, creating a web tool named AutoCompliance to simplify policy comprehension. However, this study overlooks broader GDPR coverage. In privacy policy research \cite{linden2018privacy}, the impact of GDPR on over 6,000 policies is analyzed, indicating significant revisions post-GDPR, particularly in EU policies. User experience improvements are noted, but confidence in vendor compliance remains uncertain. Leveraging ontology and text extraction techniques \cite{hu2008semantic}, vendors automate privacy policy compliance efforts, streamlining data protection measures. These advancements signify progress in managing privacy constraints, but comprehensive compliance assurance remains a challenge.

In our earlier research studies, we developed a foundational compliance knowledge graph to include various regulations and incorporated a selection of vendor privacy policy descriptions into the ontology without directly linking them to specific GDPR chapters or sections. Also, we correlated these documents with Cloud Security Alliance (CSA) controls to bridge gaps. In our past research~\cite{elluri2020measuring,elluri2018integrated}, we identified relevant sections by extracting keywords and entities from the glossaries or appendices of regulations. We then identified the semantically similar keywords
associated with GDPR regulation from the vendor privacy policies. Further, we checked for the semantic similarity between the summaries of the entire GDPR and the privacy policy document using a generic BERT abstractive summarize \cite{elluri2021bert}. In another research work, we have incorporated the National Institute of Standards and Technology (NIST) 8228 \cite{boeckl2019considerations} risk mitigation areas into the knowledge graph. \cite{echenim2023ensuring,nistir8228}. 

Expanding upon this groundwork, our current research seeks to align the extracted GDPR articles from vendor privacy policies, pinpointing any previously overlooked articles. This effort is designed to assist vendors in refining their policy documents. Given the extensive nature of these regulations, our focus is primarily on GDPR compliance, recognizing its significance for vendors handling data from EU users. The traditional approach often necessitates manual review to ensure compliance. However, our methodology proposes a more efficient solution for identifying and addressing gaps in vendor privacy policies, reducing the need for human intervention.

\subsection{Information Extraction using LLMs}
Although mostly statistical and rule-based methods have been used for information extraction \cite{pingle2019relext,piplai2020creating}, recent advancements in LLMs have opened up new methodologies.LLMs have been extensively used for information extraction across various domains. For information extraction tasks, LLMs have been utilized for generating structured entities and relationships circumventing the need to use supervised models \cite{xu2023large, peng2024metaie, li2024knowcoder}. However, to find more success in specific domains, LLMs have been fine-tuned to perform the task of information extraction. For example, in the case of scientific data extraction \cite{dagdelen2024structured} and agriculture data extraction \cite{peng2023embedding}, fine-tuned LLMs have been used. LLMs have also been used for improving annotations in the medical domain, by periodically fine-tuning based on human feedback \cite{goel2023llms}. In the domain of cybersecurity, LLMs have also been used for information combination and extraction \cite{liucyberbench,mitra2024localintel}. However, training and fine-tuning an LLM is computationally expensive and there is little guarantee that the model will not suffer from hallucinations. In our approach, we have utilized the power of RAG to limit the possibilities of hallucinations and avoid the cost of fine-tuning for our specific application scenario.

\begin{figure*}[t]
    \centering
    \includegraphics [width=\textwidth] {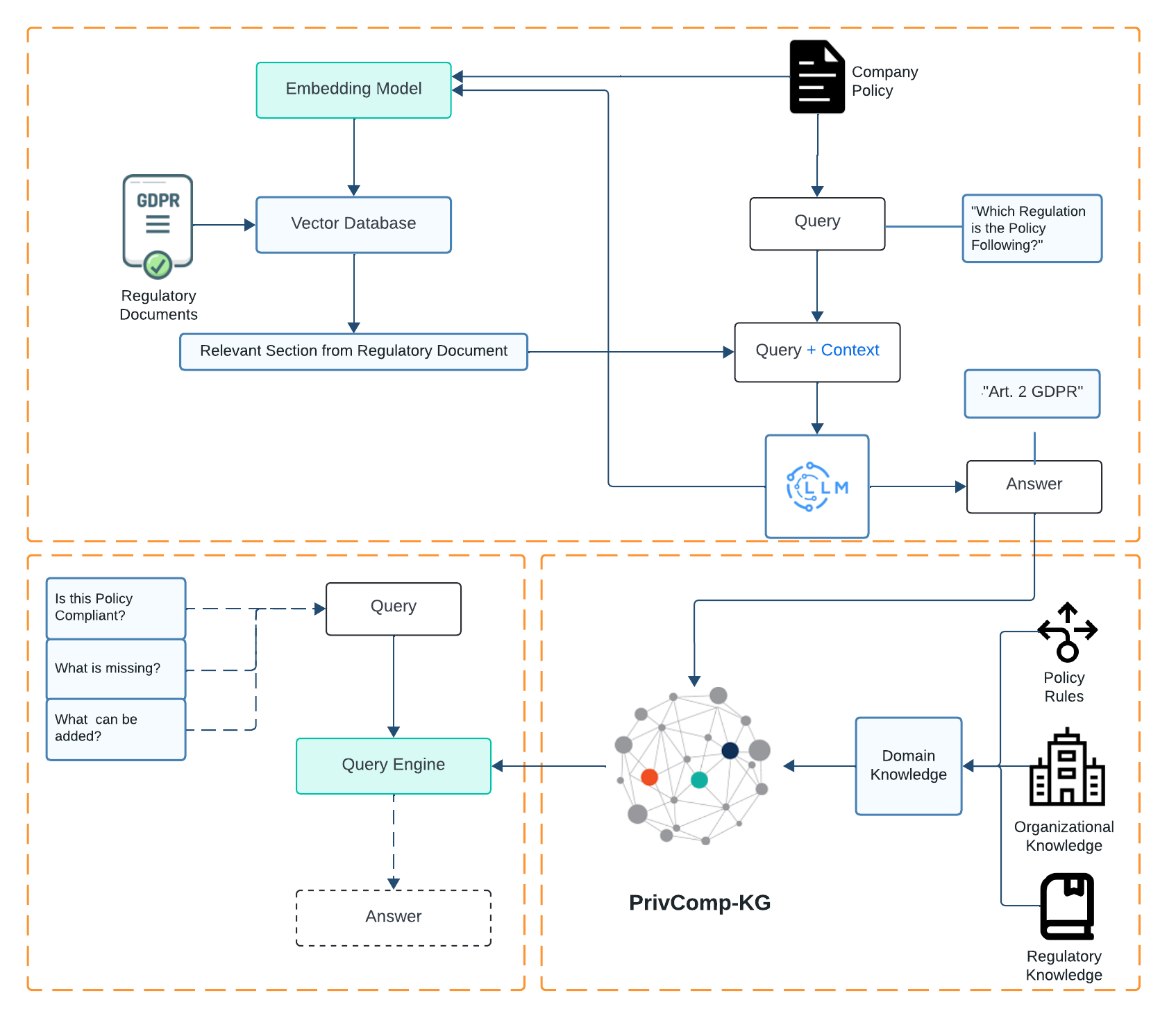} 
    \caption{Overall Framework for Building and Querying GDPR Vendor Policy Management Knowledge Graph}
    \label{fig:overall_arch}
\end{figure*}

\section{Methodology}
\label{Methodology}
The Privacy Policy Compliance Verification Knowledge Graph (\Algoname) formalizes GDPR rules and guidelines using Semantic Web technologies. It facilitates automated compliance checking, enhances transparency, and supports granular consent management. The Knowledge Graph allows for swift cross-referencing of regulatory requirements and vendor privacy policies, enabling efficient management of vendor data and adherence to regulations. Compliance inference using SWRL (Semantic Web Rule Language)  rules \cite{swrl2004semantic} enriches the understanding of privacy policies, facilitating dynamic compliance and reasoning over gaps in policies. The \Algoname is populated with relevant privacy policy properties by leveraging LLMs to assess privacy policies' alignment with regulatory articles, such as GDPR. We employ Retrieval-Augmented Generation (RAG) to mitigate LLMs' tendency for hallucinations when confronted with unfamiliar queries, dynamically generating responses without continual fine-tuning. The overall framework for \Algoname creation and population using LLMs and end user querying is demonstrated in Figure \ref{fig:overall_arch}.

\subsection{Knowledge Extraction using LLM}
\label{subsec_llm_ext}
LLMs are increasingly valued for their deep understanding of natural language. The vector representations of each piece of text written in natural language, have a deeper contextual meaning in the scope of LLMs. 

To understand privacy policies, we harness the capabilities of LLMs to assess the alignment of a privacy policy with GDPR articles. Previous efforts have focused on fine-tuning LLMs to identify similar GDPR entities corresponding to a specific privacy policy \cite{elluri2021bert}. However, both LLMs in general and fine-tuned LLMs specifically are prone to hallucinations when confronted with queries relating to unfamiliar domains. An increasingly popular method to address this issue is RAG. We opt for RAG due to the dynamic nature of the domain. Given that data privacy laws can be evolving, characterized by the emergence of new regulations, RAG aids in response generation without the need for continual model fine-tuning for every update in privacy laws and regulations.

The core components of an LLM consist of (i) a query or prompt, denoted as $P$, and (ii) a response, represented by $R$. In the context of RAG-enabled LLMs, the generation of $R$ relies on a set of documents, denoted as $D$. We utilize RAG not only to produce responses for prompts inquiring about the relationship between a privacy policy and a GDPR article but also to identify the specific segments of a GDPR article that align with a privacy policy.

\begin{figure*}[tp]
    \centering
    \includegraphics [width=13cm, height=6.25cm] {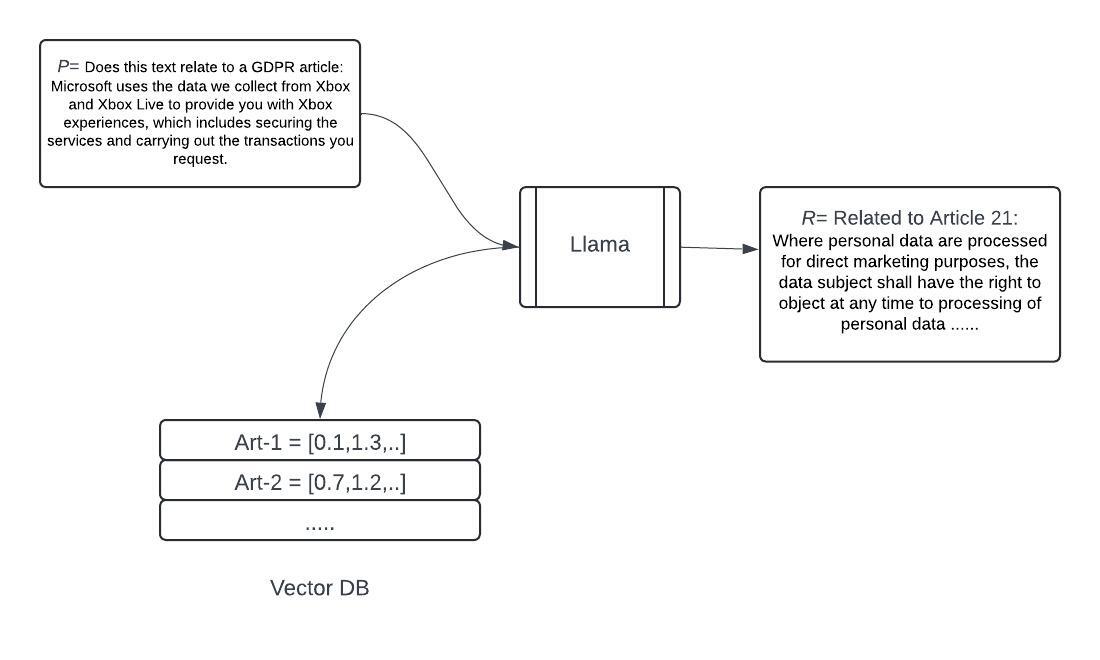} 
    \caption{Example of our model's response generation on a privacy policy sample}
    \label{fig:RAG-example}
\end{figure*}

We segment each GDPR article into chunks and integrate them into a vector store. Each segment of the GDPR article is assigned a representation within the vector store. Our vector store contains 430 chunks of GDPR articles, with each article further divided into multiple segments. Typically, each paragraph of a GDPR article corresponds to one vector within our vector store. An example of our prompt $P$ is illustrated in Figure \ref{fig:RAG-example}.

As part of the metadata for the chunks inserted into the vector store, we include the specific GDPR article from which each chunk was extracted. A typical  $P$ for this system might be "Which GDPR article does this privacy policy relate to?" followed by the privacy policy itself. The retriever then (i) generates $R$ based on the GDPR articles and (ii) provides a list of articles used to synthesize $R$, along with their corresponding similarity scores.

In Figure \ref{fig:RAG-example}, we observe an example of the model's performance using an excerpt from Microsoft's XBOX privacy policy. Our vector store is constructed by inserting chunks of GDPR articles, which are then utilized by our LLM (LLama-7B) to generate $R$. In the example, we highlight Article 21, which achieved the highest similarity score with $P$. During our experiments, we establish a threshold for the similarity score and list all articles used in generating $R$.

This process creates a comprehensive list of articles that correlate with a given privacy policy. Our system supports dynamic updates to privacy policies, as model retraining is unnecessary when policies are amended or added. The knowledge derived from privacy policies and their corresponding articles can be incorporated into our \Algoname. In the following sections, we elaborate on how the reasoning capabilities of a knowledge graph can be leveraged to derive valuable insights from privacy policies and the associated GDPR articles.

\subsection{\Algoname: Privacy
Policy Compliance Verification Knowledge Graph}
Leveraging Semantic Web technologies can streamline regulatory compliance. Using the Semantic Web for privacy policy compliance offers advantages such as standardization and machine readability. It enables automated compliance checking by representing policies in a format that software tools can understand and analyze. This approach enhances transparency by allowing policy writers to easily comprehend data privacy requirements and gaps in their existing policies. Furthermore, the semantic representations facilitate granular consent management, tailoring policies to specific regulations.

We utilized the semantic web languages Resource Description Framework (RDF) \cite{rdf} and Web Ontology Language (OWL) \cite{owl} to capture and formalize the rules and guidelines outlined in GDPR and Vendor policy documents.  We developed the novel Privacy Compliance Verification KG, \Algoname. It is designed to be in the public domain and can be adopted quickly and easily by vendors who are seeking to adhere to these regulations. The ontology is also platform-independent and can be integrated with the latest data protection regulations and many other data regulation entities. 

\begin{figure*}[tp]
    \centering
    \includegraphics [width=12.5cm, height=7.55cm] {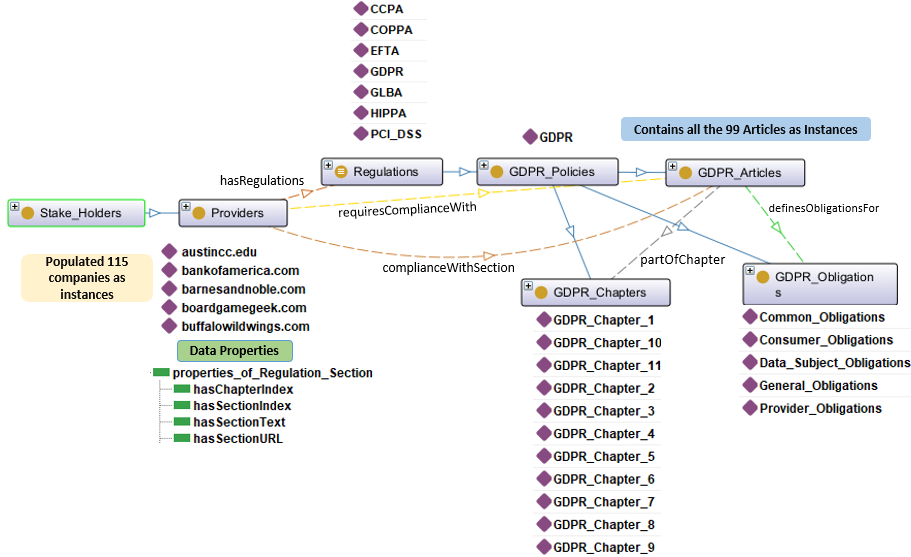} 
    \caption{\Algoname: Privacy Policy
Compliance Verification Knowledge Graph along with Instances of the classes}
    \label{fig:KG}
\end{figure*}

RDF enhances the structuring of knowledge on the web, simplifying the retrieval of domain-specific information for vendors. \Algoname is integrated with our existing Reference Document Knowledge Graph \cite{elluri2021bert}, allowing for the swift and effective cross-referencing of regulatory requirements and vendor privacy policies. As illustrated in Figure \ref{fig:KG}, this high-level knowledge graph manages GDPR rules and vendor policy extracted results. This knowledge graph is specifically designed to accommodate any applicable regulations for various vendors or companies based on the types of data they collect. Our focus on GDPR, a pivotal regulation with extensive stipulations, forms a critical component of this research methodology. Utilizing Protege software \cite{protegewiki,protege}, a free, open-source platform for ontology editing and reasoning, we constructed and managed our ontology. This structured and standardized approach not only facilitates the management of vendor data but also ensures adherence to regulations, thereby safeguarding customer data and privacy. Our knowledge graph is hosted in a public space, providing accessible user interaction.

In our earlier research, described in \cite{elluri2021bert}, we could only retrieve high-level entities and didn't examine the specific rules or articles related to privacy policies. Additionally, our previous version of the knowledge graph categorized most GDPR rules as classes, which made it challenging to compare results across different sections of the GDPR. Structuring these as instances improved our ability to identify what was missing quickly or had been extracted. We used the insights from the Section \ref{subsec_llm_ext} to update the data properties of the \textit{Provider} class. With this research, we've successfully refined our knowledge graph to manage and search GDPR rules flexibly and compare them with those extracted from vendor policy documents. We designed this knowledge graph to include a Regulations class that branches into various regulations like PCI-DSS, HIPAA, CCPA, etc. This study extracted relevant chapters, articles, and obligations from the GDPR and stored them as instances in \textit{GDPR\_Chapter}, \textit{GDPR\_Articles}, and \textit{GDPR\_Obligations}.

\subsection{Regulatory Obligations:}
\label{subsec_reg_obg}
Beyond the curation of Provider Privacy Policies and Policy Regulations, the key insight in \Algoname is drawn from the Regulatory Obligations instances. Every regulation describes based on the role of the data actor, the specific set of rules that apply to the data actor. For example, a provider, i.e. an organization offering a specific service, may collect data to support that service. However, GDPR clearly states in its provider guidelines, the specific measures that the providers need to take to protect data privacy. Additionally, some general rules with regards to the ethics and responsibility of data collection apply to the providers. This knowledge about the specific role and application of GDPR sections is encoded in the \textit{GDPR\_Obligations} class. Specifically, there are 5 instances of the \textit{GDPR\_Obligations}, each describing a specific role:
\begin{itemize}
    \item \textit{Consumer\_Obligations} Consumers must inform the supervisory authority and the data subject about any personal data breaches. Additionally, the consumer must conduct a Data Protection Impact Assessment (DPIA), consult with the supervisory authority before processing if the DPIA indicates a high risk, and appoint a Data Protection Officer when processing personal data on a large scale.
    \item \textit{Common\_Obligations} Rules that apply to both consumers and providers, who are responsible for ensuring compliance.
    \item \textit{Data\_Subject\_Obligations} Consumers must inform data subjects about the duration for which, or why, data will be retained upon collection. Consequently, if data subjects request the removal of their data, and it is no longer necessary for the purposes for which it was collected, it must be erased.
    \item \textit{General\_Obligations} GDPR mentions generic rules that are not specific to any role but apply to all the data processing activities in general.
    \item \textit{Provider\_Obligations} Provider is primarily responsible for assisting consumers in the event of data breaches and processing data in accordance with consumer directives. Additionally, the Provider must maintain comprehensive records of all data processing activities and ensure robust data security measures are in place to protect consumer information.
\end{itemize}

\subsection{Compliance and Regulatory Properties:}
The \Algoname supports object properties that links classes to support compliance reasoning and verification.
\begin{itemize}
\item{hasRegulation:} Regulations identified in vendor privacy policies are stored as instances within the \textit{Regulation} class and linked to the \textit{Providers} class.

\item{compliesWithSection:} Using the knowledge extraction tool, as described in Section \ref{subsec_llm_ext}, the regulatory articles that individual provider privacy policies complies with are identified. This knowledge is populated into the \Algoname using the "compliesWithSection" relation between \textit{Providers} 
and \newline \textit{GDPR\_Articles}. 

\item{requriesComplianceWith:} Every vendor handling EU user data must adhere to relevant GDPR requirements. To enforce this, we link the \textit{Providers} class to the \textit{GDPR\_Chapters} through this object property.

\item{partOfChapter:} Since a chapter can encompass multiple articles, we connect the \textit{GDPR\_Articles} to the \textit{GDPR\_Chapters} class to reflect this relationship.
\end{itemize}

Additionally, to organize the findings from privacy policy analyses, we established several data properties:
\begin{itemize}
\item{hasChapterIndex:} This property stores the indices of chapters identified in a policy document.
\item{hasSectionIndex, hasSectionText, and hasSectionURL:} These properties are essential for recording the sections extracted from the documents, including descriptions and URLs for easy reference.
\end{itemize}

\subsection{Compliance Inference}
SWRL rules enhance reasoning capabilities in Semantic Web applications by allowing the specification of logical rules that define relationships and infer new information from existing data. The \Algoname supports SWRL rules that infers necessary rules from regulatory articles based on the role of the data actor and the role based obligations, as described in Section \ref{subsec_reg_obg}. 

For example, for the privacy policy of a provider that wishes to collect and utilise data, the following SWRL rules can infer the obligatory rules from GDPR.

\begingroup
\makeatletter
\@totalleftmargin=-1cm
\begin{verbatim}
    S1: Cloud_Providers(?cloud_provider) ^ GDPR_Articles(?gdpr_article) 
    ^ definesObligationsFor(?gdpr_article, Provider_Obligations) 
    -> requiresComplianceWith(?cloud_provider, ?gdpr_article)
\end{verbatim}

\begin{verbatim}
    S2: Cloud_Providers(?cloud_provider) ^ GDPR_Articles(?gdpr_article) 
    ^ definesObligationsFor(?gdpr_article, Common_Obligations) 
    -> requiresComplianceWith(?cloud_provider, ?gdpr_article)
\end{verbatim}
\endgroup

By applying these SWRL rules, \Algoname can derive detect inconsistencies in existing policies, and make logical deductions about compliance. This reasoning process enriches the understanding and interpretation of privacy policy data, facilitating a dynamic approach towards privacy policy compliance, more advanced semantic querying and data integration. After processing with the reasoner, users can efficiently compare the required rules against those extracted, updating any lacking areas in the privacy policy to ensure it is current and compliant.

\section{Experimental Results}
\label{exp_res}

\subsection{Dataset}
The OPP-115 dataset \cite{wilson2016creation} is a comprehensive collection of privacy policies from various online platforms, consisting of over 115 privacy policies. It encompasses a wide range of websites and services, including social media platforms, e-commerce sites, and mobile applications. The dataset is structured and annotated, making it suitable for privacy policy analysis and evaluating language models. It includes the categorisation of data collection, usage, sharing, and retention practices outlined in the privacy policies. In this work, we use the OPP-115 dataset to demonstrate the utility of our model and evaluate the performance of the LLMs. Our LLM based method is used to identify relevant regulatory articles for each provider in the dataset and then populated into \Algoname. The inferential engine in \Algoname reasons over the gaps in these provider policies. The results are made available to end users using a query engine. The results from our evaluation methods and query engine are described in subsequent sections. 

\begin{figure*}[t]
    \centering
    \includegraphics [width=0.8\textwidth] {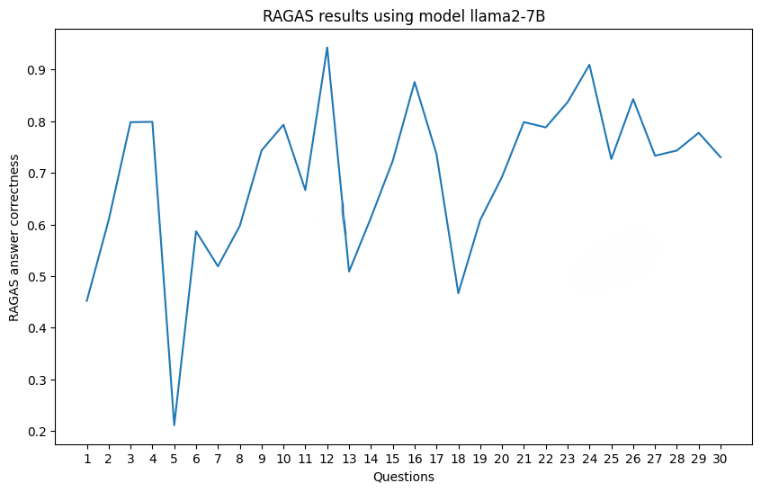} 
    \caption{RAGAS \cite{es2023ragas} score evaluating the quality of privacy policy "Answer" generation through our LLM, in comparison with human-generated "Answers". A score of `1' means perfect alignment.}
    \label{fig:ragas}
\end{figure*}

\subsection{Evaluation of LLM-guided extraction}

\begin{wraptable}{R}{6cm}
\centering
\resizebox{0.45\textwidth}{!}{%
\begin{tabular}{|l|l|}
\hline
Threshold Used & Correctness Score \\ \hline
0.9            & 0.66              \\ \hline
1.0            & 0.74              \\ \hline
1.1            & 0.82              \\ \hline
1.2            & 0.84              \\ \hline
1.3            & 0.89              \\ \hline
1.4            & 0.88              \\ \hline
1.5            & 0.9               \\ \hline
\end{tabular}%
}
\caption{Correctness score for each threshold}
\label{tab:correctness}
\end{wraptable}

For our experiments, we use Llama-7B LLM, and \textit{chromaDB} as our vector store. The knowledge extracted from the RAG-enabled LLM focuses on GDPR articles corresponding to specific portions of a privacy policy. We evaluate this knowledge in two ways: (i) assessing the quality of RAG-LLM generated responses and (ii) determining the accuracy of the retrieved article numbers by the LLM. 

As RAG is integral to the LLM, we assess the effectiveness of $R$ generated by our model given $P$. A common metric for evaluating RAG-generated responses is RAGAS \cite{es2023ragas}. Our model's responses are compared with those generated by humans. Figure \ref{fig:ragas} illustrates the correctness scores for all questions. Questions with low scores (<0.5) required more specific information regarding the privacy policies. Since our LLM only accesses GDPR articles, it performs well in questions regarding the similarity of a privacy policy to a GDPR article but struggles with overly specific questions about the policy itself.

The second metric utilizes the OPP-15 dataset. Since the OPP-15 dataset maps privacy policies to GDPR articles through an intermediate category, we leverage it to assess the accuracy of the articles retrieved during $R$ generation. We pre-process the dataset to eliminate portions of a privacy policy that are too small. RAG offers a similarity score function, which we apply a threshold to, selecting only the chunks that fall below a certain threshold. Table \ref{tab:correctness} presents the accuracy of correctness for various thresholds.

\begin{figure}[t]

\begin{subfigure}{0.5\textwidth}
  \centering
  \includegraphics[width=\textwidth]{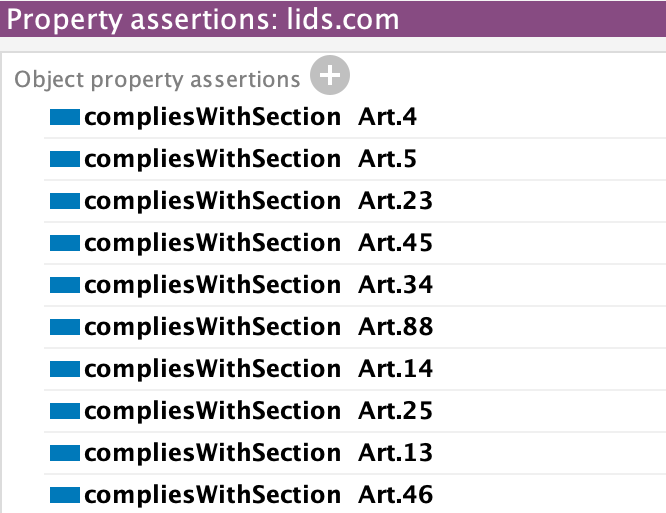}
  \label{fig:swrl_1}
  \subcaption[]{GDPR articles with which the privacy policy currently complies}
\end{subfigure}
\begin{subfigure}{0.5\textwidth}
\centering
  \includegraphics[width=\textwidth]{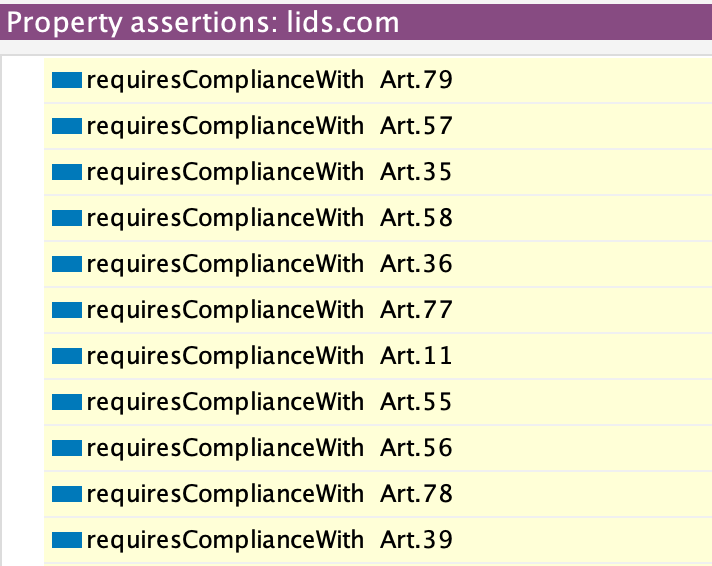}
  \label{fig:swrl_2}
\subcaption[]{GDPR articles with which the privacy policy needs to comply}
\end{subfigure}
\caption{Compliance results for Lids.com from \Algoname}
\label{fig:SWRl_result}
\end{figure}%

\subsection{Knowledge Graph Inferences and Queries}
The \Algoname is populated with the provider privacy policies from the OPP-115 dataset. Furthermore, the identified regulatory sections of relevance to the privacy policies are populated with "compliesWithSection" object property. Finally, using the reasoner, \Algoname automatically updates the GDPR articles that are mandatory for the given provider using the "requiresComplianceWith" object property. 

Fig \ref{fig:SWRl_result} demonstrates the results for the provider "Lids.com". Leveraging the LLM results, the \Algoname is populated with the GDPR articles that "Lids.com" has compliance with, as demonstrated in Fig \ref{fig:SWRl_result} (a). Using the KG reasoner, \Algoname also populates the GDPR articles that require compliance for "Lids.com", as demonstrated in Fig \ref{fig:SWRl_result} (b). These information are critical to policy writers of providers, like "Lids.com", in identifying gaps in their existing policies. To query the missing articles for a provider that need to be addressed, we provide the following SPARQL query:

\begin{minipage}{20em}
\texttt{ SELECT * WHERE {
{cc:lids.com cc:requiresComplianceWith ?x}
MINUS {cc:lids.com cc:compliesWithSection ?x}
}}
\end{minipage} \\

For "Lids.com", the \Algoname return 40 articles that has been identified as mandatory according to GDPR, but missing from the existing privacy policy. This information helps easily identify the gaps in the privacy policy and effectively identify the guidelines that need to be addressed.

\section{Conclusion and Future Work}
In the digital age, safeguarding data protection and privacy has become paramount. To effectively manage their operations, many companies enlist the support of third-party vendors and service providers. These partners play critical roles in various aspects of business functions, such as data handling, storage, and processing. Entrusting sensitive data to third parties necessitates a rigorous approach to ensure comprehensive data protection and privacy. Companies must establish robust protocols and contractual agreements to safeguard the integrity and confidentiality of the information shared with these external entities. Most existing research works focus only on specific sections of the regulation, which won't be helpful for organizations. In this research, we leverage LLM and Semantic Web technologies to verify compliance of privacy policy documents with policy regulations, like GDPR. This work also proposed the novel Privacy Policy Compliance Verification Knowledge Graph, \Algoname, for storing and retrieving comprehensive information to maintain privacy policies. This proposed research enhances the readability of privacy policies and also promotes transparency, empowers consumers, strengthens regulatory compliance, and ultimately fosters trust in the digital ecosystem. We aim to further assist companies by enriching our \Algoname with the requisite U.S. federal and state-level data protection regulations, facilitating rapid cross-referencing across this comprehensive legislative framework.

\subsubsection{\discintname}
Authors have no competing interests.
\subsubsection{Supplemental Material Statement:}
\begin{itemize}
\item{The \Algoname is available from: 
\newline
https://github.com/ 
<anonauthor>/\Algoname.git} 
\item{Datasets are available from: https://www.usableprivacy.org/data \cite{wilson2016creation}}
%\item{Query sets are available from:}
\item{Source code for populating \Algoname can be accessed from: 
\newline
https://github.com/
<anonauthor>/\Algoname.git}
\end{itemize}

\bibliographystyle{splncs04}
\bibliography{ref}

\end{document}